\documentclass[aps,pra,showpacs,twocolumn,preprintnumbers,amsmath,amssymb,superscriptaddress]{revtex4}

\usepackage[dvips]{graphics,color}
\usepackage{amsmath}
\usepackage{amsfonts}

\usepackage{amssymb}
\usepackage{amstext}

\usepackage{graphicx}
\usepackage{dcolumn}
\usepackage{bm}
\usepackage{bbm}

\newcommand{\ket}[1]{|#1\rangle}

\newcommand{\id}{\mathbbm{1}}

\begin{document}
\newcommand{\ignore}[1]{}

\title{Random circuits by measurements on weighted graph states}

\author{A. Douglas K. Plato}
\email{alexander.plato@imperial.ac.uk}
\address{Institute for Mathematical Sciences, Imperial College London, Prince's Gate, London SW7 2PG, UK\\QOLS, Blackett Laboratory, Imperial College London, Prince Consort Road, London SW7 2BW, UK}

\author{Oscar C. Dahlsten}
\email{dahlsten@phys.ethz.ch}
\address{Institute for Theoretical Physics, ETH Zurich, 8093 Zurich, Switzerland}
\author{Martin B. Plenio}
\email{m.plenio@imperial.ac.uk}
\address{Institute for Mathematical Sciences, Imperial College London, Prince's Gate, London SW7 2PG, UK\\QOLS, Blackett Laboratory, Imperial College London, Prince Consort Road, London SW7 2BW, UK}

\date{\today}

\begin{abstract}
Random quantum circuits take an input quantum state and randomize it. This is a task with
a growing number of identified uses in quantum information processing. We
suggest a scheme to implement random circuits in a weighted graph state. The input state is entangled with the weighted graph state and a random circuit is implemented when the experimenter performs local measurements in one fixed basis only. The scheme uses no classical random numbers and is a new and natural application 
of weighted graph states.
\end{abstract}

\pacs{03.67.Bg, 03.67.Mn, 05.40.-a}
\maketitle
\section{Introduction}
Randomising inputs has a wide usage in information processing. In the case of quantum information processing, the input may correspond to a pure state and the randomisation a unitary evolution picked at random. The random unitary cannot be generated directly, but instead needs to be implemented as a sequence of random elementary gates, i.e. a random circuit.

The number of applications for random circuits in quantum information process is growing. They may be applied to hide information about input states \cite{DiVincenzoLT02}, to sample state spaces \cite{EmersonAZ05}, and to model random processes such as thermalisation \cite{SerafiniDP06, SerafiniDP07, CalsamigliaHDB05, GemmerOM01,PopescuSW06,OliveiraDP06,DahlstenOP07} and black hole information leakage \cite{HaydenP07}. A more surprising application is in the superdense coding of quantum states \cite{HarrowHL04}. Generating uniformly random operators over $U(2^N)$ is, however, exponentially hard. Instead it is often sufficient to produce circuits which are identical in only a few of the relevant statistical properties. Such pseudo-random circuits have already been efficiently implemented in an $NMR$-setup \cite{EmersonLL03}.

In the present work we propose a new experimental implementation for such random circuits. One prepares, in advance, a highly entangled resource state called a weighted graph state. The input state is entangled with a number of qubits in the resource state and the experimenter then performs local measurements in one fixed basis only. The randomness of the measurement outcomes effectively pick the circuit, and the output state is then carried by a subset of qubits of the total system.

We identify several advantages of our scheme. No classical random numbers are used in the process, so the randomness is entirely quantum. Furthermore, the measurement based approach has the advantage of being comparatively scalable. The fact that the experimenter only needs to measure in one basis throughout is a simplification and is to our knowledge at present unique to our scheme. The scheme is furthermore a new application of weighted graph states which can help to motivate such experiments.

Recent work has independently suggested the measurement based approach for generating random circuits \cite{BrownWV09}. They were able to improve on a previously known gate model, by first translating it into the conventional cluster state formalism. Here, however, we find that using weighted graph states is arguably more natural and simple: no classical pseudo-random numbers are required, and measurements are only performed in one basis. 

We proceed as follows. Firstly background material on random circuits, measurement based quantum computation and weighted graph states is presented. We then define and analyse our proposed method. In the subsequent sections the scheme is evaluated, and finally we summarize possible experimental implementations and discuss open questions.

\section{Background}

\subsection{Random Circuits}
In this section we define random quantum circuits and then list properties we deem desirable, motivated by the consideration of possible applications.

A sequence of quantum gates, i.e. unitaries, is called a circuit. Here {\em a sequence of quantum gates picked at random} is called a {\em random circuit}. Thus according to this definition, a simple example of a random circuit for a single qubit is the following: apply the Hadamard\footnote{$H: \ket{0/1} \mapsto \ket{+/-}$.} gate $H$, Pauli $X$ or $\id$ each with probability $1/3$. However, in practice, we are often more interested in random circuits with particular properties, and so we include three additional conditions that such proposals should satisfy. These will serve as the basis for our later evaluation. 

{\em (i)Unbiased sampling asymptotically}\newline
For hiding an input state fully one requires the output states to have a flat distribution in state space, i.e. according to the unitarily invariant Haar measure $P$ where $P(\ket{\Psi})=P(U\ket{\Psi})$. It is a simple but important observation that if the gates are picked from a non-universal set one would not access all states, and therefore not achieve the uniform distribution, even in the asymptotic limit. In fact it is essentially sufficient for a random circuit to consist of gates picked from a universal set of gates for it to induce the Haar measure asymptotically - see \cite{EmersonLL05} for discussion.  

{\em (ii) Good sampling in polynomial time}\newline
Unbiased sampling of the uniform distribution to within a fixed accuracy requires $exp(N)$ elementary gates, where $N$ is the number of qubits \cite{EmersonLL05}. 
However for practical applications one will require the circuit to give a sufficiently good sampling in a 
modest, i.e. $poly(N)$ time. What merits being called a sufficiently good sampling can depend on the application.
For example, in superdense coding of quantum states \cite{HarrowHL04, DahlstenOP07, DahlstenPhD} one requires that the entanglement is typically maximal \cite{HaydenLW06,Lubkin78,LloydP88,Page93,FoongK94}.  In fact, typically maximal entanglement may be achieved to some accuracy within $poly(N)$ elementary gates picked at random. Such schemes exist \cite{OliveiraDP06, DahlstenOP07}, results which were recently generalised in \cite{HarrowL08}.

{\em (iii) Feasible experimental implementation}\newline
For practical applications we furthermore require a feasible experimental implementation, ideally one that is possible with current technology. 

It should be noted that the unbiased sampling of property {\em (i)} is often included in definitions of random (and pseudo-random) circuits found elsewhere in the literature. 

\subsection{Measurement based quantum computation}
In this section we briefly introduce one-way, also called measurement based, quantum computing together with some relevant notation. Firstly we describe how to evolve a state by measurements followed by an introduction to the idea of resource states for measurement based quantum computation. For a more comprehensive introduction we refer the reader to \cite{HeinDERVB06}. 

A quantum computation is an evolution of an input state to an output state. The unitary transformation can in fact be realized through measurements. For example, say that one wants to perform the Hadamard gate, $H$, on a single qubit in state $\ket{\Psi}=\alpha\ket{0}+\beta\ket{1}$. We start by first preparing a second qubit in state $\ket{+}=\frac{1}{\sqrt{2}}\left(\ket{0}+\ket{1}\right)$. Then we perform a Control-Z gate (CZ)\footnote{CZ: $\ket{00}\mapsto\ket{00},\ket{01}\mapsto\ket{01},\ket{10}\mapsto\ket{10},\ket{11}\mapsto -\ket{11}$.} on the two qubits, yielding the state  $CZ(\alpha\ket{0}+\beta\ket{1})\ket{+}=\alpha\ket{0}\ket{+}+\beta\ket{1}\ket{-}$. Finally, we make a projective measurement in the $\ket{+}/\ket{-}$ basis on the {\em first} qubit. Now, if an outcome of $+1$ is obtained, the new state will be $\ket{+}(\alpha\ket{+}+\beta\ket{-})$. Thus the {\em second} qubit is now in the state $H\ket{\Psi}$, as desired. One notes that there is a random choice of the gate applied, depending on the two possible measurement outcomes. For controlled computations this randomness is in general a problem, albeit a surmountable one \cite{HeinDERVB06}. However, for the purpose of producing random circuits this effect can, on the contrary, be an advantage.
 
\begin{figure}
\centering \includegraphics[width=8.0cm]{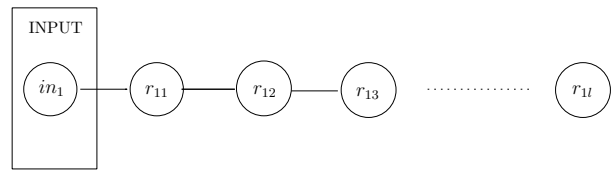}
\caption{A resource state for quantum computation on a single qubit together with the input state. Circles are qubits and the qubit ($in_1$) carries the input state. The figure shows the state before the measurements 
have been performed. Solid lines indicate $CZ$ gates. The resource qubits are in the $\ket{+}$ state before the $CZ$ is applied. They are labelled $r_{11},r_{12}...r_{1l}$, where $l$ signifies the length of the circuit and the first and second index is the row and column in the graph respectively.}
\label{fig:n1resourcestate}
\end{figure}

In order to perform a sequence of gates one can reiterate this procedure, i.e. introduce a third qubit in state $\ket{+}$, apply CZ on qubits 2 and 3 and so forth. However the CZ operations commute with the relevant measurements in such a way that one can equivalently perform the CZ gates first, before doing any measurements. All the non-input qubits can be prepared in this manner and be thought of as a {\em resource state}, to be used together with an arbitrary input state. Figure \ref{fig:n1resourcestate} describes a possible resource state for the case of one qubit. This can naturally be generalised to more qubits, which will be discussed in the next section.  

It is often argued that the resource state measurement-based approach is easier to implement in experiment than the circuit model because one can do most of the entangling operations before the input state has been given. Since these gates are particularly difficult to do, it is desirable to be able to fail and redo them, without losing the input state. Only when this has been successfully achieved does one give the input state. This requires entangling the qubits carrying that state with those in the resource. After which, only local measurements on the system are necessary to achieve an arbitrary unitary transformation \cite{HeinDERVB06}.

\subsection{Weighted graph states}
There are many types of resource states one can use for measurement based quantum computation. In this section we define a type of resource state called a graph state \cite{HeinDERVB06}.  Then we introduce a natural generalization thereof called weighted graph states, which is what we will use for our scheme.  

We begin by defining a \textsl{graph}, $G=(V,E)$, as a finite non-empty point set $V$ along with a collection $E\subset V$ of unordered pairs of points in $V$. We say $V$ is the collection of the points or vertices of $G$ and $E$ the collection of edges. A graph is \textsl{simple} if it is undirected, and has no loops or weights. We can describe a graph by the elements of a matrix $\Gamma_{ab}$, where entries correspond to whether an edge connects vertices $a$ and $b$. For unweighted graphs, $\Gamma_{ab}=1$ for all $\{a,b\}\in E$ and zero otherwise. Such a matrix is called an adjacency matrix of the graph $G$, and so a simple graph can be described by a symmetric matrix, of ones and zeros, having zeros along the main diagonal.  

Graphs such as these can be used to describe a family of quantum states in the following manner. We first consider each vertex as labeling a qubit in the state $|+\rangle$. Between each pair of qubits connected by an edge in the associated graph we apply a unitary operation $U_{ab}(\Gamma)=e^{-i\Gamma_{ab}\frac{\pi}{4}(I-\sigma_z^a)\otimes(I-\sigma_z^b)}$, i.e. a CZ gate. The resulting states described by this procedure are graph states, and includes the cluster state resources as special cases. Using this formulation, it is then straightforward to generalise graph states to weighted graph states (WGS), by simply relaxing the condition that edges carry no weights. Thus, the simplest definition of a weighted graph state is,
\begin{equation}\label{eq:wgsdef}
|\Psi_{WGS}\rangle = \prod_{\{a,b\}\in E} U_{ab}(\Gamma_{ab})|+\rangle ^{\otimes N},
\end{equation}
where the product is taken over all edges $\{a,b\}$ and the unitaries are now defined by $U_{ab}=e^{-i\Gamma_{ab} \frac{\pi}{4}(I-\sigma_z^a)\otimes(I-\sigma_z^b)}$, where $\Gamma_{ab}$ are given by the components of the adjacency matrix of a weighted graph. 

Various applications and generalisations of weighted graph states themselves have been proposed \cite{AndersPDVB06,HartmannCDB07,GrossE07}. For example, one can include more general initial states, additional filtering operations and local unitary operators. However, for the purposes of this paper, the above definition captures the relevant ideas needed to extend the cluster state formalism. 
\begin{figure}
\centering \includegraphics[width=8.0cm]{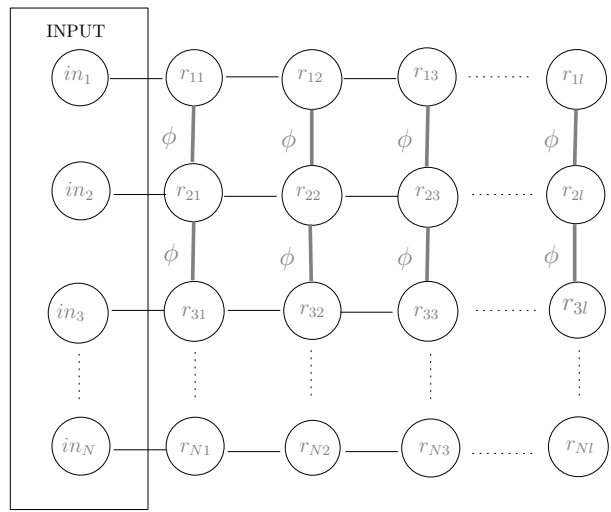}
\caption{The initial weighted graph state resource used in our random circuit
generation scheme prior to any measurements, together with the column of qubits carrying the input state. Circles represent qubits and the qubits ($in_1, ...in_N$) carry the input state.  Thin horizontal lines indicate $CZ$ gates, while thicker vertical lines are $\phi$ gates, defined in equation \ref{eq:phigate}. The resource qubits are in the $\ket{+}$ state before the $CZ$ and $\phi$ gates are applied. Each lie on a vertex $a=({j,k})$ in a graph $G$ and take the labeling $r_{jk}$. $j$ is the row index which runs from 1 to $N$, the number of qubits in the input state. $k$ is the column index which runs from $1$ to $l$,  the length of the circuit.}
\label{fig:wgs}
\end{figure}
%
\section{Measurement based scheme}\label{sec:mbs}

We now describe our scheme for generating random circuits, using ideas from measurement based quantum computation. The key feature in this proposal is the adoption of weighted graph states as the resource states for measurement based quantum computing. This enables us to present a simple, fixed measurement algorithm which does not have the requirement of a separate classical (or quantum) random number generator. It follows closely the simplest cluster state model, with only minor variations in the non-local operations made between adjacent rows.

First, we describe the proposed scheme. Then we characterise the associated evolution of the input state. This is important for showing that the process is unitary.

The proposal is as follows. First, a planar rectangular array of $N\times l$ qubits is prepared, each initially in the state $\ket{\psi}_{r_{jk}}=\ket{+}$ with subscripts $j$ and $k$ labeling the row and column positions along the array, see Fig \ref{fig:wgs}. This configuration is represented by a graph $G$ embedded in a $2$D lattice, and so a vertex, $a$, is described by two numbers $(j,k)$, and on each vertex lies a qubit $r_{jk}$. Between each pair of adjacent qubits on neighbouring columns we apply $CZ$ gates, that is, $U_{r_{jk}r_{jk\pm1}}=\mathbf{diag}(1,1,1,-1)$. Between each pair of adjacent qubits on neighbouring rows we apply the control phase gates,
\begin{equation} 
\label{eq:phigate}
U(\phi_{r_{jk}r_{j\pm1k}})=\mathbf{diag}(1,1,1,e^{-i\phi_{r_{jk}r_{j\pm1k}}}), 
\end{equation}
where the parameters $\phi_{ab}=\Gamma_{ab}\frac{\pi}{4}=\phi$ for all $a$, $b$. We will refer to this as the $\phi$-gate. As these operators commute, they can be carried out in parallel. The number of rows needed is determined by the number of qubits on which the random unitary will act, while the depth or length $l$ specifies the number of iterations to be performed. Note that it will not be necessary to prepare the entire resource state to depth $l$ initially - it may be grown as the protocol proceeds. The circuit acts on an $N$ qubit state $\ket{\Psi}_{in}$ carried by qubits $in_1$, $in_2$ ... $in_N$. A $CZ$ gate is applied between each $in_j$ and $r_{j1}$ qubit. Then each $in$ qubit is measured in the basis $\{\ket{+},\ket{-}\}$. The possible outcomes are labeled by a bi-vector $\vec S^{(k)}$ where each component $S^{(k)}_j$ takes the value $S^{(k)}_j=0$ for a measurement yielding the eigenvalue $1$ and $S^{(k)}_j=1$ for the eigenvalue $-1$.

The measurements on the input column projects the first column of resource qubits into the state,
\begin{equation}\label{eq:evoltimestep}
\ket{\psi}_{r_{j1}}^{\otimes_j}=G^{(1)}M(\vec S^{(1)})\ket{\Psi}_{in}
\end{equation}
where the operators acting horizontally between rows are given by
\begin{equation}
M(\vec S^{(k)})= M_1(S^{(k)}_1)\otimes M_2(S^{(k)}_2) \otimes \cdots \otimes M_n(S^{(k)}_n),
\end{equation}
with
\begin{equation}
M_j(S^{(k)}_j) = H Z^{S^{(k)}_j},
\end{equation}
and the vertical operators by
\begin{equation}
G^{(k)} = \prod_{\{(j,k),(j',k)\}\in E} U(\phi_{r_{jk}r_{j'k}}).
\end{equation}
Where, again, the set of edges $E$ is represented in Fig. \ref{fig:wgs}. Measurements are then performed successively on columns $1$ through $l-1$, leaving the output state on the final line of qubits. Immediately, we can see that the evolution from column to column is unitary, and so the entire process can be described by a single unitary operation. In fact, as there is no feed-forward of the measurement outcomes one could in principle make all measurements simultaneously.

In Appendix \ref{app:gts} we consider more general weighted graph states and more general measurements. However, as the next section will show, the simpler version of the scheme appears to work very well.

\section{Evaluation of scheme}

\subsection{Unbiased sampling asymptotically}
We demonstrate numerically that the scheme gives unbiased sampling (Haar measure) in the asymptotic time limit. More precisely we show that the output state distribution passes a necessary and stringent test, namely that the entanglement probability distribution approach that associated with the Haar distribution on pure states. 

The entanglement distribution associated with the Haar distribution has received considerable attention recently \cite{HaydenLW06,Lubkin78, LloydP88, Page93, FoongK94}. The average entropy of entanglement $\mathbbm{E} S_A(N_A,N_B)$ of a set of $N_A$ spins was studied already in the 70s and 80s \cite{Lubkin78,LloydP88} and the explicit solution(`Page's conjecture') was conjectured in \cite{Page93} and proven in \cite{FoongK94}:
\begin{equation}
\label{eq:page}
\mathbbm{E} S_A(N_A,N_B)=\frac{1}{\ln2}\sum_{k=2^{N_B}+1}^{2^{N_A+N_B}} \left( \frac{1}{k}-\frac{2^{N_A}-1}{2^{N_B+1}}\right)
\end{equation}
with the convention that $N_A \leq N_B$ and where $N_A+N_B=N$, the total number of particles.

This can be used to show that the average entanglement is very nearly  maximal, meaning close to $N_A$, for large quantum systems, i.e. $N\gg 1$. Hence one concludes that a randomly chosen state will be nearly maximally entangled with a large probability. Indeed, it was recently shown that the probability that a randomly chosen state will have an entanglement $S_A$ that deviates by more than $\delta$ from the mean value $\mathbbm{E} S_A(N_A,N_B)$ decreases exponentially with $\delta^2$ \cite{HaydenLW06}.

Achieving this entanglement probability distribution is a strong condition to claim one has the true Haar distribution. Often one quantifies multipartite entanglement using ``linearised'' measures based on purity. For example, a popular measure used for pseudo-random circuits is the Meyer-Wallach entanglement \cite{MeyerW02}, which is related to the average purity of each qubit. However, stabilizer states give exactly the correct purity one would expect from the Haar measure. Nevertheless they do not yield the correct entanglement distribution \cite{SmithL05,DahlstenP06}. The probability distribution of entanglement, $P(S_A)$ associated with stabilizer states sampled uniformly at random is given by Theorem I in \cite{DahlstenP06},
\begin{eqnarray}\label{eq:stabst}
P(S_A)=\frac{\prod_{i=1}^{N_A}(2^i+1)}
{\prod_{k=N-N_A+1}^{N}(2^k+1)}\cdot \nonumber\\
\cdot\prod_{j=1}^{S_A}
\frac{ \left(2^{N-N_A+1-j}-1\right)\left( 2^{N_A+j}-
2^{2j-1}\right)}{2^{2j}-1 }
\end{eqnarray}
where $N_A$ is the number of qubits belonging to Alice and $N$ is the total number of qubits. Again $N_A \leq N-N_A=N_B$. The total state is bipartite and pure.

\begin{figure}
\centering \includegraphics[width=8.0cm]{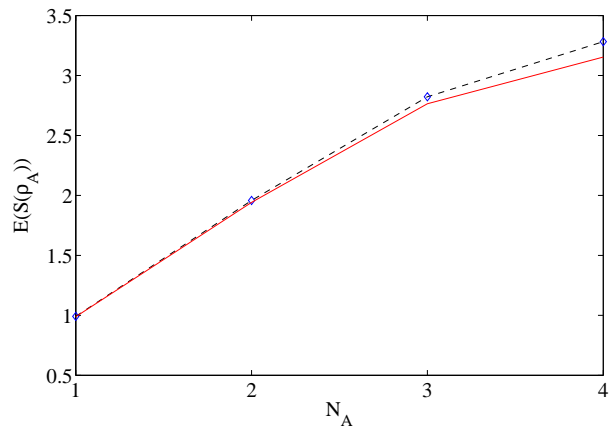}
\caption{(Color online) The average entanglement, $\mathbbm{E}(S(\rho_A))$, (blue diamonds) of our randomized circuits compared with the average expected entanglement given by Page's conjecture (dashed line) for $N=8$ and various $N_A$. The solid red line shows the same quantities for stabilizer gates chosen at random. To calculate our averages we first perform $10^6$ iterations on a randomly chosen initial state. Then we average the entanglement calculated over the next $10^6$ iterations. In each case the fractional difference is less than $10^4$ from that expected from true random circuits.}
\label{fig:pagetest}
\end{figure}

\begin{figure}
\centering \includegraphics[width=7.0cm]{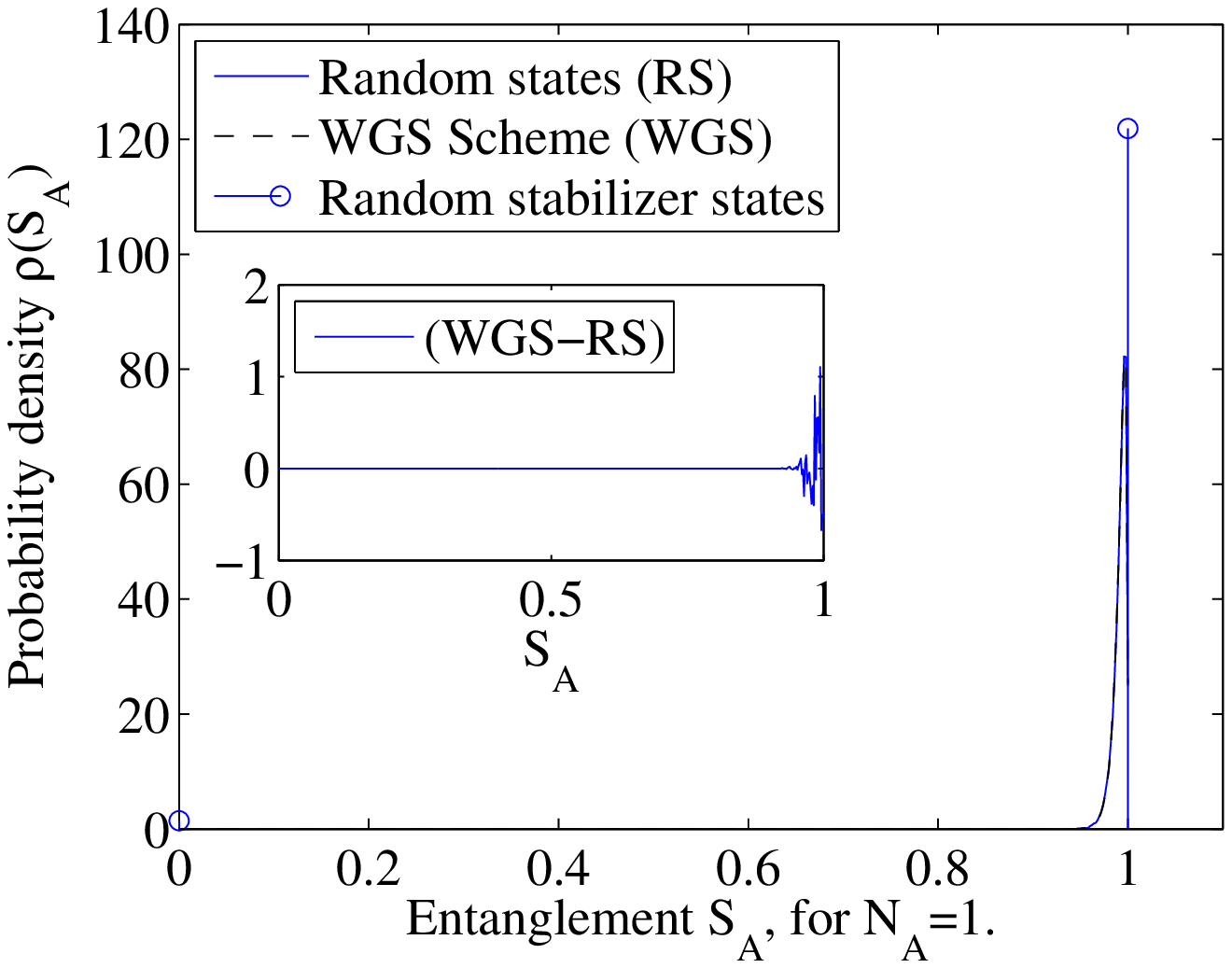}
\includegraphics[width=7.0cm]{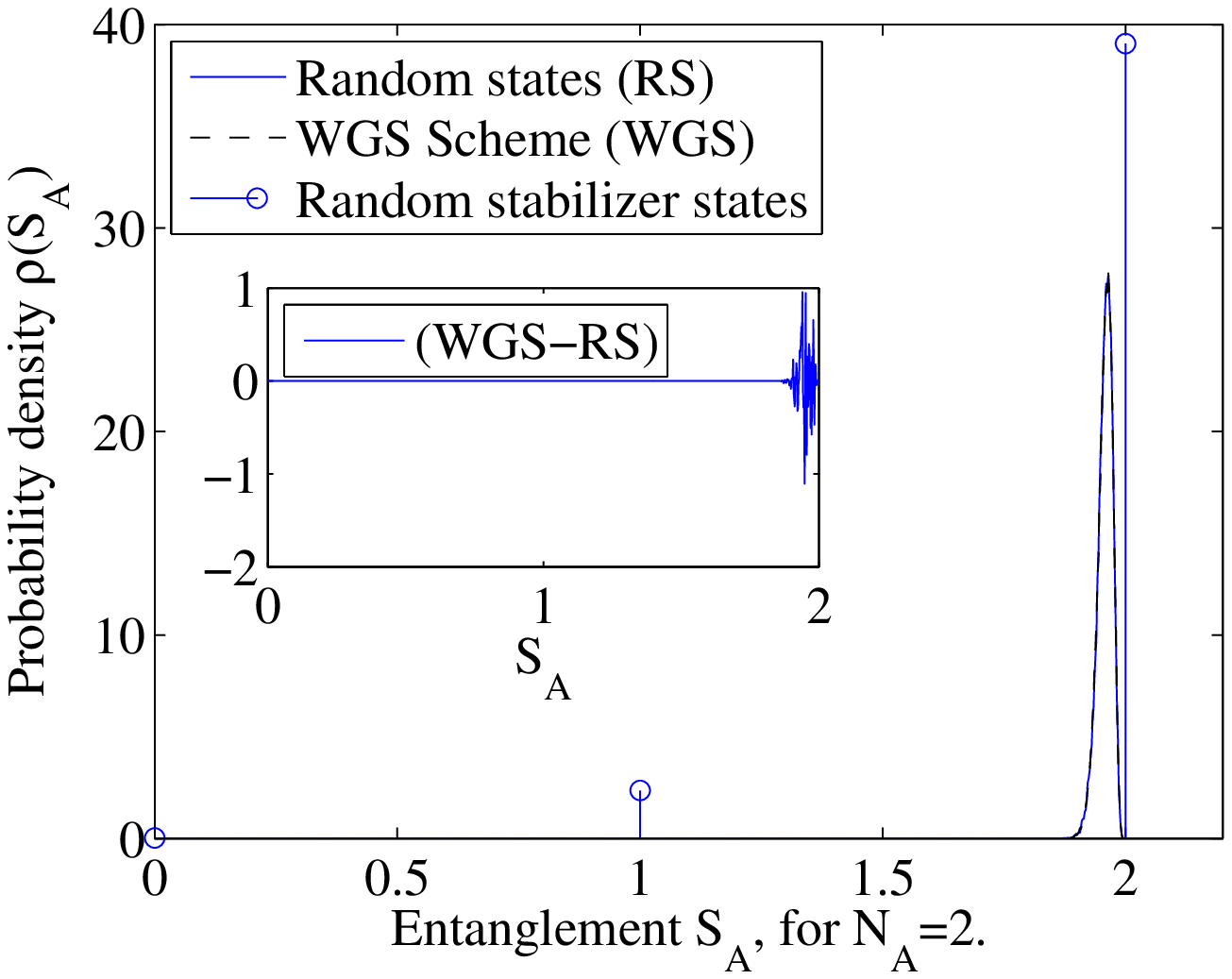}
\includegraphics[width=7.0cm]{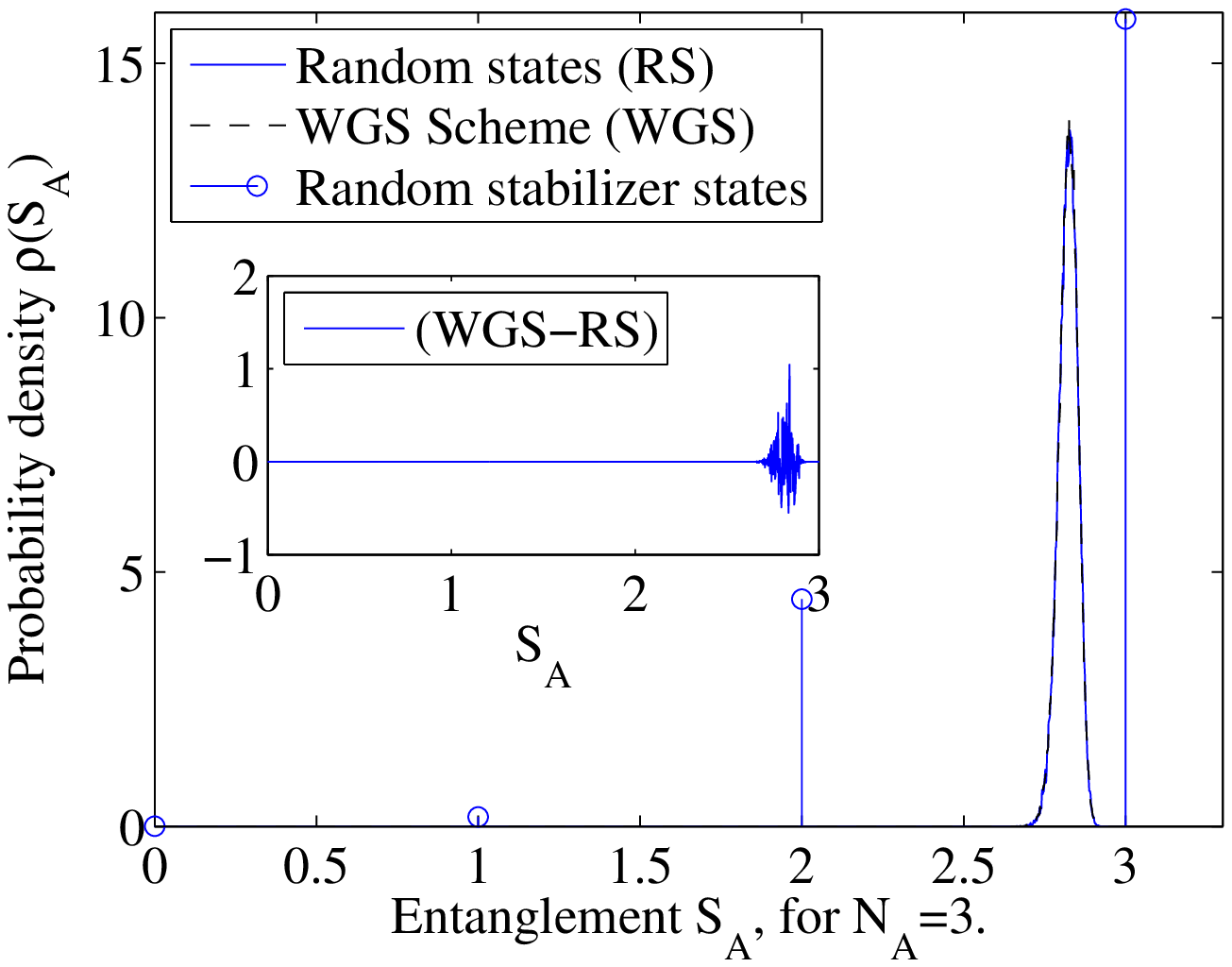}
\includegraphics[width=7.0cm]{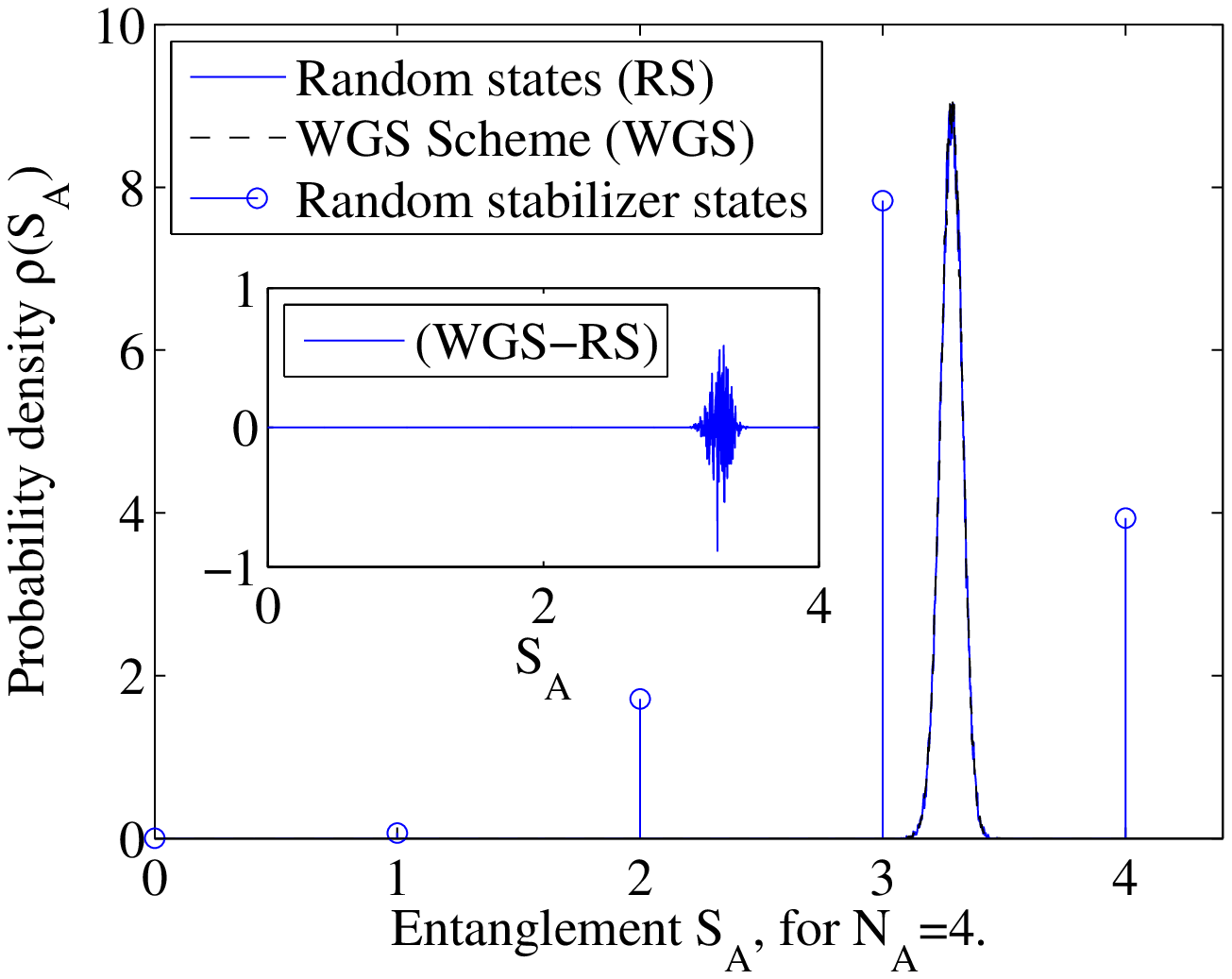}
\caption{(Color online) Comparisons of the entanglement distributions between the weighted graph states and states chosen at random with respect to the Haar measure, for $N=8$ and $N_A=1,2,3,4$ (plots from top to bottom). In both cases we calculate $10^5$ samples and group into $500$ evenly spaced bins. The distributions are then re-scaled such that $\int \rho(S_A) = 1$. Black dashed lines represent the weighted graph state entanglements, and blue solid lines represent the Haar distribution. For comparison, we also include the distributions for stabilizer states chosen at random \ref{eq:stabst} (solid lines capped with circles). }
\label{fig:pagetest2}
\end{figure}

Figures \ref{fig:pagetest} and \ref{fig:pagetest2} show that the output states do indeed follow the correct entanglement statistics. We also add the relevant stabilizer statistics for comparison, which highlight the fact that averaging over stabilizer states is not as good in this regard.   

The above considerations strongly indicate that the states coming out of the scheme in the asymptotic time limit are indeed Haar distributed, but they do not constitute an analytical proof\footnote{One could attempt a proof along the lines of \cite{DeutschBE95}, by showing that the operators effected by the various measurement outcomes can be strung together to generate any linear combination of elements spanning the Lie Algebra $u(2^N)$. However, we have now restricted ourselves to specific one-parameter families of unitary operators, and so can no longer invoke the same generic arguments conventionally applied in such approaches. This leads to many difficulties. For example, when trying to demonstrate the appropriate irrationality conditions for the eigenvalues of the evolution operators explicitly, the resulting analytic expressions quickly become intractable. Thus, we have so far been unable to prove that fixed measurements and weighted graph state resources generate a universal set. It should be mentioned, however, that one can immediately see why fixed measurements on graph state resources are not universal. This would correspond to a parameter value of $\phi=\pi$ in our model, for which the eigenvalues of the resulting unitary operators each contain one of the pairs $\pm \pi/3$ or $\pm 2\pi/3$. As these operators are periodic, we can not generate arbitrary linear combinations of their algebra elements, and so we can never cover the full space of unitary operations. }. 

\subsection{Good sampling in poly(N) time}
In the previous section we considered the sampling in the infinite time limit. In practise one will have a finite time available, and this time should scale as a polynomial function in $N$, the number of qubits carrying the input state. Under this restriction one cannot generate the Haar measure exactly, however one can hope to obtain a sufficiently good sampling of the state space nevertheless. 

\begin{figure}
\centering \includegraphics[width=8.0cm]{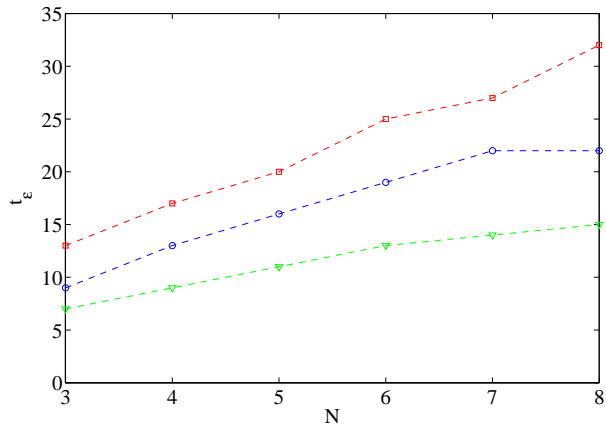}
\caption{(Color online) The scaling with $N$ of the time taken for the average entanglement to agree with that associated with the Haar measure to a fixed accuracy $\varepsilon<0.0001$ (red squares), $\varepsilon<0.001$ (blue circles) and $\varepsilon<0.01$ (green triangles). For $N=2,3,4$ averages are over $10^6$ realisations. For $N=5,6$ averages are over $10^5$ realisations and for $N=7,8$ averages are over $10^4$ realisations. In all cases $\phi=5\pi/8$.}
\label{fig:pagetest3}
\end{figure}

To test this for our current scheme, we again use the entanglement of the output states and compare it with the expected entanglement of states chosen at random with respect to the Haar measure. We demand that the difference between the two expected entanglement values is consistently less than some small $\varepsilon$. This will occur after some time $t_{\varepsilon}$, for a given $N$ and $N_A$. We then check how $t_{\varepsilon}$ scales with increasing $N$ with fixed $N_A$, Fig. \ref{fig:pagetest3}.

These results suggest that the time required to achieve the Haar average is linear in the number of qubits and so we expect good scalability of our scheme\footnote{Note that the `time' here refers to the number of columns in the weighted graph state that have been measured. Care should therefore be taken in comparing it with the number of two qubit gates necessary in a circuit model randomization. Note that for each time step here, $2N-1$ two-qubit gates are needed, which effectively performs $N-1$ two-qubit gates on the input state, even though only local measurements are made. The experimenter however needs to perform successfully $(2N-1)l$ two qubit gates in total, where $l$ is the length of the circuit to be implemented.}. It is worth noting here the relevance of the parameter $\phi$. Numerical indications suggest this is related to the absolute rate of convergence, with a maximum obtained by a value of approximately $\phi=5\pi/8$.  It can be easily seen that two cases for which our scheme does not work are $\phi=\pi$ and $\phi=2\pi$. These lead, respectively, to $2$D and $1$D cluster state resources, for which fixed measurements are not universal. Interestingly, however, we have seen no evidence of any other general restrictions on permissible values of $\phi$. An intuitive explanation for this is that the set of possible single time-step evolutions, eq. (\ref{eq:evoltimestep}), are non-commuting operators. Taken together these may explore the space of unitaries with some complicated dependence on the parameter $\phi$. 

%

\subsection{Realization}
There are several methods and technologies in which states such as the one presented in Fig. \ref{fig:wgs} may be realized. Here we briefly mention two. Firstly, in optical lattices it is possible to implement two-body collisions that implement a Hamiltonian of the form $H = J \sigma_z^{(1)}\otimes \sigma_z^{(2)}$. To obtain the state in Fig. \ref{fig:wgs} it is then necessary and possible to first implement this Hamiltonian between horizontal neighbors for a time $t = \pi/(4J)$ followed by the implementation of the same pairwise Hamiltonian between vertical pairs for a time $t=\phi/(2J)$ \cite{Mandel}. The ability to generate the desired quantum state in parallel is contrasted by the slight disadvantage that local addressability in such systems is difficult. Recently proposed spin and polariton systems in arrays of optical cavities \cite{HartmannBP06} allows, in principle, for mechanisms analogous to the one described above \cite{HartmannBP07} but where individual nodes of the cluster state reside in distinct cavities and are therefore addressable. 

A somewhat different approach to achieve states such as those in Fig. \ref{fig:wgs} is via measurement induced interaction between cavities. The basic idea relies on the insight that one may first entangle the electronic degree of freedom of atoms inside two distinct cavities with the cavity photons. These photons will then leak out of the cavities where they are mixed on a $50/50$ beamsplitter and then detected at the two outputs of the beam-splitter. This detection may implement a Bell projection on the atoms enabling the generation of entangled states \cite{Bose}. Such Bell projections can be used to build large cluster states. These Bell projections may then actually be used directly to implement Controlled NOT gates \cite{Protsenko} in a loss tolerant way. Weighted graph states can then be generated via the application of controlled phase gates with rotation angle $\phi$. These gates may be obtained from two controlled NOT gates supplemented by local rotations \cite{9authors}.

\section{Conclusion}

We have proposed a scheme to generate random quantum circuits. In our method, an experimenter prepares a particular type of very entangled state called a weighted graph state. The qubits carrying the input state are then entangled with part of the weighted graph state and the experimenter then just performs projective measurements in the $\ket{+}/\ket{-}$ basis. The randomness of the measurement outcomes chooses the particular circuit and no classical random numbers are necessary. We have tested the effectiveness of this scheme in randomising the input state by comparing the entanglement distribution of the output to that associated with the uniform distribution on states. Numerical results strongly suggest these two are identical, and so we may be conclude that the scheme is indeed very effective.

From the theoretical perspective several important and interesting questions appear. Firstly one should prove rigorously that the scheme can generate any unitary operation. Furthermore, it may also be viewed be an interesting setting to study thermalization, since there is no classical randomness inserted by hand, yet the evolution appears to maximise the entropy of the input state.

\begin{acknowledgements}
The authors would like to thank Koenraad Audenaert, Oliver Butterley, Dan Browne, David Gross, Jens Eisert, Roberto Oliveira and Renato Renner for useful comments and discussions. ADKP would also like to thank Renato Renner for hospitality at ETH Zurich, during the write up of much of this work. ADKP and MBP were supported by the EPSRC QIP-IRC GR/S82176/0). MBP is also supported by the EU Integrated Project QAP funded by the IST directorate under contract number 015848 and is supported by a Royal Society Wolfson
Research Merit Award. OCD was supported by the Imperial College Institute for Mathematical Sciences and the Swiss Federal Institute of Technology

\end{acknowledgements}


\newpage
\begin{appendix}

\section{Generalising the scheme}\label{app:gts}
One could also consider generalising the scheme presented in section \ref{sec:mbs}, by including resource qubits in the state $\ket{\psi}_{r_{jk}}=\gamma_{jk}\ket{0}+ \delta_{jk}\ket{1}$, along with general two qubit operations between columns. Furthermore, we could also allow measurements in the basis $\{ ( |0\rangle \pm e^{i\theta_{jk}} |1\rangle)/\sqrt 2) \}$ . Defining new operators $\overline{U}_L^{(s)}=\langle0|_a U_{ab}|0\rangle_a
+ (-1)^s e^{-i\theta_{jk}}\langle1|_a U_{ab}|0\rangle_a$ and $\overline{U}_R^{(s)}=\langle0|_a U_{ab}|1\rangle_a + (-1)^s e^{-i\theta_{jk}}\langle1|_a U_{ab}|1\rangle_a$, $a=r_{jk}$, $b=r_{jk+1}$ we can write
the more general evolution operators as
\begin{widetext}
\begin{eqnarray*}
M_j(\vec S^{(k)})=
\left(\!\!\!
\begin{array}{cc}
\gamma\langle0|_b \overline{U}_{L}^{S^{(k)}_j}|0\rangle_b \! + \! \delta\langle0|_b \overline{U}_{L}^{S^{(k)}_j}\!|1\rangle_b
&\gamma\langle0|_b \overline{U}_{R}^{S^{(k)}_j}\!|0\rangle_b \! + \! \delta\langle0|_b \overline{U}_{R}^{S^{(k)}_j}\!|1\rangle_b \\
\gamma\langle1|_b \overline{U}_{L}^{S^{(k)}_j}\!|0\rangle_b \! + \! \delta\langle1|_b \overline{U}_{L}^{S^{(k)}_j}\!|1\rangle_b
&\gamma\langle1|_b \overline{U}_{R}^{S^{(k)}_j}\!|0\rangle_b \! + \! \delta\langle1|_b \overline{U}_{R}^{S^{(k)}_j}\!|1\rangle_b
\end{array}
\!\!\!\right).
\end{eqnarray*}
\end{widetext}
For operators $U$ with entries $u_{ij}$ this can be written as,
\begin{equation}
M_j(\vec S^{k})= 
\left(
\begin{array}{cc}
m_{11}
&m_{12} \\
m_{21}
&m_{22}\end{array}
\right).
\end{equation}
where
\begin{eqnarray*}
m_{11}\! &=& \!\gamma ( u_{11}+(-1)^{S^{(k)}_j} e^{-i\theta_{jk}}u_{31})\!  \nonumber \\
&& \qquad\qquad\qquad\qquad + \; \delta ( u_{12}+(-1)^{S^{(k)}_j}e^{-i\theta_{jk}}u_{32}), \nonumber \\
m_{12} \! &=& \!\gamma ( u_{13}+(-1)^{S^{(k)}_j}e^{-i\theta_{jk}}u_{33})\! \nonumber \\
&& \qquad\qquad\qquad\qquad + \; \delta ( u_{14}+(-1)^{S^{(k)}_j}e^{-i\theta_{jk}}u_{34}), \nonumber \\
m_{21}\! &=& \!\gamma ( u_{21}+(-1)^{S^{(k)}_j}e^{-i\theta_{jk}}u_{41}) \! \nonumber \\
&& \qquad\qquad\qquad\qquad + \; \delta ( u_{22}+(-1)^{S^{(k)}_j}e^{-i\theta_{jk}}u_{42}), \nonumber \\
m_{22}\! &=& \! \gamma ( u_{23}+(-1)^{S^{(k)}_j}e^{-i\theta_{jk}}u_{43}) \! \nonumber \\
&& \qquad\qquad\qquad\qquad + \; \delta ( u_{24}+(-1)^{S^{(k)}_j}e^{-i\theta_{jk}}u_{44}).
\end{eqnarray*}
If $U$ is a diagonal unitary operator, then we have the following restriction for unitary evolution,
\begin{equation}
|\gamma|^2 u_{11}^* u_{33} +|\delta|^2 u_{22}^* u_{44}=0.
\end{equation}
For some applications, such as hiding information about an input state, we are interested in obtaining a randomising operation itself, as opposed to just a state with an expected amount of entanglement. This operation should ideally be unitary, and so the simplest guarantee is to demand that each step in the circuit also corresponds to a unitary operation. We then must impose the above condition from the outset, and so if we wish to adhere to the weighted graph state formalism (\ref{eq:wgsdef}), then we are forced to apply only CZ gates between the neighbouring columns of Fig. \ref{fig:wgs}.

\end{appendix}

\newpage 
\bibliography{apssamp}

\end{document}